# Deep ATLAS radio observations of the CDFS-SWIRE field

Ray P. Norris[1], José Afonso[5], Phil N. Appleton[2], Brian J. Boyle[1], Paolo Ciliegi[6], Scott M. Croom[4], Minh T. Huynh[2], Carole A. Jackson[1], Anton M. Koekemoer[7], Carol J. Lonsdale[3], Enno Middelberg[1], Bahram Mobasher[7], Seb J. Oliver[8], Mari Polletta[3], Brian D. Siana[2], Ian Smail[9], Maxim A. Voronkov[1]

1: CSIRO Australia Telescope, CSIRO Radiophysics Laboratory, PO Box 76, Epping, NSW 1710, Australia. Ray.Norris@csiro.au
2: Spitzer Science Center, MS220-6, California Institute of Technology, Pasadena, CA 91125, USA
3: Center for Astrophysics and Space Sciences, University of California, San Diego, 9500 Gilman Drive, La Jolla, CA 92093-0424, USA.
4: Anglo-Australian Observatory, PO Box 296, Epping, NSW 1710, Australia.
5: Observatório Astronómico de Lisboa, Faculdade de Ciências, Universidade de Lisboa, Tapada da Ajuda, 1349-018 Lisbon, Portugal, & Centro de Astronomia e Astrofísica da Universidade de Lisboa, Portugal.
6: INAF, Osservatorio Astronomico di Bologna, Via Ranzani 1, I-40127, Bologna, Italy.
7: Space Telescope Science Institute, 3700 San Martin Drive, Baltimore, MD 21218, USA
8: Astronomy Centre, CPES, University of Sussex, Falmer, Brighton BN19QJ, UK
9: Institute for Computational Cosmology, Durham University, South Road, Durham DH1 3LE, UK

## Abstract

We present the first results from the Australia Telescope Large Area Survey (ATLAS), which consist of deep radio observations of a 3.7 square degree field surrounding the Chandra Deep Field South, largely coincident with the infrared Spitzer Wide-Area Extragalactic (SWIRE) Survey. We also list cross-identifications to infrared and optical photometry data from SWIRE, and ground-based optical spectroscopy. A total of 784 radio components are identified, corresponding to 726 distinct radio sources, nearly all of which are identified with SWIRE sources. Of the radio sources with measured redshifts, most lie in the redshift range 0.5-2, and include both star-forming galaxies and active galactic nuclei (AGN). We identify a rare population of infrared-faint radio sources which are bright at radio wavelengths but are not seen in the available optical, infrared, or X-ray data. Such rare classes of sources can only be discovered in wide, deep surveys such as this.

## 1. Introduction

Large multiwavelength surveys in the last few years have proved to be powerful tools for understanding galaxy formation and evolution, particularly those that use obscuration-independent tracers of star-formation and AGN activity, such as radio, mid-infrared, and far-infrared wavelengths. However, surveys at these wavelengths have typically covered only small areas, and so not only suffer from cosmic variance, but are also likely to miss intrinsically unusual objects. Those that have covered wider areas have been relatively shallow, and so may have missed the most active epochs of galaxy formation.

The Spitzer Wide-area Infrared Extragalactic Survey (SWIRE) program (Lonsdale et al. 2003) has addressed these limitations by observing large (nearly fifty square degrees in total) fields at mid- and far-infrared wavelengths with sufficient sensitivity to detect highly-obscured, ultraluminous infrared galaxies (ULIRGs) at $z>>1$. SWIRE's goal is to trace the evolution of dusty, star-forming galaxies, evolved stellar populations, and Active Galactic Nuclei (AGN), from redshifts $z\sim3$, when the Universe was 2 Gyr old, to the present day.

Over the last two years, we have conducted the Australia Telescope Large Area Survey (ATLAS) of the Chandra Deep Field South (CDFS) and European Large Area ISO Survey - South 1 (ELAIS-S1) regions, with the aim of producing the widest (6 square degrees) deep (10-15 μJy rms) radio survey ever attempted. The surveyed areas have been chosen to overlap the SWIRE areas as far as practicable, so that infrared and optical data are available for most of the radio objects. They also encompass the well-studied Great Observatories Origins Deep Survey (GOODS) field in the CDFS (Giavalisco et al., 2004).

The broad scientific goals of this multi-wavelength survey are to understand the formation and evolution of galaxies in the early Universe. The radio observations are complementary to the Spitzer observations in being able to detect radio AGN in even the most obscured galaxies, and provide additional information on the spectral energy distribution (SED) of the galaxies. Galaxies powered by star formation are expected to follow the radio-far infrared correlation, whilst departure from this correlation is a strong indicator of an Active Galactic Nucleus (AGN).

Surveys of radio sources with flux densities greater than 1 mJy are typically dominated by AGN, but source count statistics suggest the presence at sub-mJy levels of another population (Condon, 1984; Windhorst et al. 1985; Hopkins et al. 2003), which has been attributed to star forming galaxies (e.g. Windhorst et al. 1985; Georgakakis et al. 1999, Afonso et al. 2005). However,

Chapman et al. (2003a) show that many of these weaker galaxies have relatively high redshifts (z>1) and luminosities ($L_{20cm} > 10^{23}$ $WHz^{-1}$). It is not clear whether this increased luminosity is caused by abnormally high star-formation rates, such as those found in ULIRGs, or is the result of an AGN, possibly embedded within a star forming galaxy.

There have been a number of very important deep radio surveys (e.g. Condon et al., 2003; Hopkins et al. 2003) which have produced valuable data on radio source statistics, but the potential power of these surveys is often hampered by inadequate data at other wavelengths. The ATLAS survey is specifically targeted on wide areas which are, or will be, well-studied at other wavelengths. It is thus uniquely capable of testing the alternative hypotheses, as we have a large number of galaxies with extensive radio, infrared, and optical data (and in some cases deep X-ray data), and the ATLAS survey should prove pivotal to understanding these objects.

Radio AGNs in the local Universe are typically divided into radio-loud objects (e.g. radio quasars, radio galaxies), whose radio luminosity is generally > $10^{24}$ $WHz^{-1}$, and radio-quiet objects (e.g. Seyferts and normal galaxies). It is not yet clear whether this classification is relevant to the early Universe, where we find, for example, double and triple radio sources which resemble classical radio-loud sources, but whose radio luminosity is significantly lower. Unlike the well-studied objects in the local Universe, we do not yet understand the evolution of radio sources in the early Universe. For example, Magorrian et al. (1998) have shown that in the local Universe the mass of the super-massive black hole (SMBH) in a galaxy is related to that of the bulge of the galaxy. We do not know whether this is true in the early Universe, nor how it is related to the star formation rate. Particularly interesting are those cases where the radio source lies buried within a host galaxy whose optical/infrared spectrum or SED appears to be that of a star-forming galaxy. Understanding the relationship between the AGN activity and the star-forming activity in these galaxies is a primary goal of this project.

Obscured activity may be the dominant contributor to galaxy luminosity at high redshifts, and hence any purely optically-derived model of galaxy formation is likely to be incomplete. Spitzer surveys have shown that the Cosmic Infrared Background is dominated by luminous infrared galaxies at around z~1 (Dole et al., 2006), while Chapman et al. (2003b) suggest that a large population of highly-obscured, but very active galaxies at z=1-5 may be the dominant location of massive star formation and AGN-fuelling at high redshifts. Moreover, there is evidence that these galaxies are strongly clustered, and also have a correlation-length exceeding any other known high-z population (Blain et al. 2004; Stevens et al. 2003). These results suggest that these dusty, high-redshift galaxies trace the growth of large scale structure in the early Universe, and are the precursors of the massive galaxies in the local Universe.

The specific science goals of ATLAS are:

- To test whether the radio-far-infrared (FIR) correlation changes with redshift or with other galaxy properties. Once calibrated, this correlation, which is thought to be driven by active star formation, will be a powerful tool for determining the star formation history of the Universe.

- To search for over-densities of high-z ULIRGs which mark the positions of proto-clusters in the early Universe. With a sampling volume of $2 \times 10^7$ $Mpc^3deg^{-2}$ (in the range z = 1-3) this survey will contain at least one proto-cluster with a present-day mass equivalent to Coma. There are expected to be tens of lower-mass systems undergoing the first phase of their collapse in this volume, all of which can be detected from the tracer population of obscured ULIRGs which are thought to reside in such regions (Stevens et al. 2003).

- To trace the radio luminosity function to a high (z~1) redshift for moderate-power sources, and measure for the first time the differential 20cm source count to a flux density limit of ~30 μJy to a high precision.

- To open a region of parameter space, corresponding to a large area of sky surveyed with high sensitivity at radio, mid-infrared, and far-infrared wavelengths, which would enable us to discover rare but important objects, such as short-lived phases in galaxy evolution.

We are currently about half-way through the ATLAS survey observations, having covered 6 square degrees of the CDFS and ELAIS-S1 fields to a sensitivity of about 40 μJy. When the survey is complete, we hope to reach a final rms of 10-15 μJy (depending on time allocation) over this field, and will then release data products including FITS images and source catalogues. We also plan to observe the field at another radio wavelength to obtain spectral indices, to obtain complementary ground-based optical spectroscopy on the radio sources, and conduct Very Long Baseline Interferometry (VLBI) observations of a subset of sources.

In this paper, we present the results obtained from the data taken in 2004 on the CDFS-SWIRE field, to provide a first look at the stronger radio sources in this field.

Throughout this paper, we define a radio "component" as a region of radio emission identified in the source extraction process. We define a radio "source" as being one or more radio components which appear to be physically connected, and which probably correspond to one galaxy. Thus we count a classical triple radio-loud source as being a radio source consisting of three radio components, but count a pair of interacting starburst galaxies as being two sources, each with one radio component.

Throughout this paper, we use the following cosmological parameters: $H_0$ = 71 $\mathrm{kms^{-1}Mpc^{-1}}$, $\Omega_m$=0.27 and $\Omega_\Lambda$=0.73.

# 2. Observations and Data Reduction

## *2.1 Radio observations and data reduction*

The first radio observations of the region surrounding the GOODS field in the CDFS (AT Project C1035) were taken by Koekemoer et al. (2003, 2006), to which we refer hereafter as KAMNC, and have been compared with optical and X-ray data by Afonso et al. (2006). The data were observed over a mosaic of seven overlapping fields, which are shown as dashed lines in Fig. 1. A total of 149 hours of integration were used by KAMNC, or 21.3 hours per pointing.

The ATLAS survey observations (AT Project C1241) cover a much wider area, which was chosen to cover both the CDFS and ELAIS-S1 SWIRE fields. Here we concentrate solely on the SWIRE-CDFS region, and all figures here and throughout this paper refer solely to these CDFS observations. The CDFS area was covered by a mosaic of 21 pointing centers shown in Fig.1. The first observations took place in January 2004 and are expected to continue until 2007. The data used in this paper are those taken up to the end of 2004, and include a total of 173 hours of integration, or 8.2 hours per pointing.

The observations were taken in the AT mosaic mode, in which the array was cycled around the 21 pointing centers, spending about 2 minutes on each, together with observations of the secondary calibrator 0237-233 at least once per cycle. Observations were taken in a variety of array configurations to maximize u-v coverage (i.e. the sampling of the Fourier plane). The shortest baseline was 31 m, and the longest was 6000 m. The observing dates and configurations used in this paper are shown in Table 1.

In this paper, we use data from both C1035 and C1241. Because noise and sidelobes from our shorter observations extend into the area observed by KAMNC, we do not achieve as high sensitivity in this region as them, and we have not taken any special steps to do so. Instead, this paper should be regarded as complementary to the results of KAMNC, who observe a small area with high sensitivity, whereas in this paper we observe a large area with lower sensitivity.

To avoid the regions at the edge of the field which have significant primary beam attenuation, we restrict the area covered in this paper to the RA range $03^h26^m$ to $03^h36^m$, and the declination range -29º00' to -27º 12' giving a total surveyed area of about 3.7 square degrees.

The CDFS-ATLAS field contains an unusually strong source (S145 = ATCDFS_J032836.53-284156.0) in pointing centre 1, which presents a challenge to our calibration procedures, as it is present in the sidelobes of several other pointings. Calibration errors from this source significantly increase the rms noise of the images in this region of the ATLAS field. We have found that a significant contributor to these calibration errors is the non-circularity of the primary beam response of the antennas. While the primary beam response may be measured accurately, current radio-astronomy imaging packages do not enable the data to be corrected for this. Work is in progress both to characterize the primary beam response (using holographic antenna measurements) and to write new calibration software that can apply this information.

All observations were made with two 128 MHz bands, centred on 1344 and 1472 MHz. The correlator was used in continuum mode (2 x 128 MHz bandwidth), with each 128 MHz band divided into 32 x 4 MHz channels, and all four Stokes parameters were measured.

The primary flux density calibrator used was PKS B1934-638, which is the standard calibrator for ATCA observations (S = 14.95 Jy at 1.380 GHz; Reynolds 1994). We calibrated the complex antenna gains by frequently (typically every 20–40 minutes, depending on atmospheric phase stability) observing the secondary calibrator PKS 0237-233. The resulting phase errors are typically at the level of a few degrees before self-calibration and are not a significant limiting factor in the resulting images.

We used the Australia Telescope National Facility (ATNF) release of the MIRIAD (Sault et al. 1995) software to reduce our data. Before imaging, the data from each observing session were inspected and the MIRIAD interactive tasks *tvflag* and *blflag* were used to flag bad data resulting from interference or hardware problems. The primary calibrator data were flagged before calibration was applied. The secondary calibrator and target data had bandpass and polarization calibration applied before inspection and flagging.

The radio-frequency interference environment at Narrabri deteriorated significantly during the course of the observations: at the start of the observations described here, only minimal flagging was required, whereas by late 2004 about 30% of the data in the second IF had to be deleted because of interference. The data were first flagged using an automated system based on cross-polarization products, and were then manually flagged by inspecting rms, amplitude, and phase on each baseline as a function of time.

When imaging, we explored a range of weighting schemes, and eventually chose a super-uniform weighting scheme which yielded high spatial resolution, but at the expense of sensitivity. Thus the rms noise of the images used here is about 50% higher than could be obtained by using natural weighting. For example, KAMNC used robust weighting, with a robustness parameter of 0.5, and thus reached a significantly lower flux density, but at poorer spatial resolution.

Because of the large observing bandwidth (2 x 128 MHz), the multi-frequency synthesis (Sault & Wieringa 1994) technique was necessary to improve u-v coverage and reduce bandwidth smearing. This technique makes a single image from multi-frequency data by gridding each spectral channel in its correct place in the u-v plane, instead of at a location determined by the average over all channels.

Bandpasses for each day were calibrated on 1934-638 using *mfcal*, and the resulting bandpass tables copied to other observations on the same day. The first stage of the processing was to image field 1, which contains the strong 1 Jy source ATCDFS_J032836.53-284156.0, and then perform four iterations of selfcal on this source (two with phase only, and the last two with both phase and amplitude). The gain solutions from this selfcal process were then copied to observations of other fields on the same day. Further selfcal on other fields was also tried, but was not found to improve calibration significantly.

The "individual" approach to mosaicing was taken, in which each pointing field was imaged (using a multi-frequency synthesis) and cleaned separately, and then a linear mosaic (using *linmos*) used to mosaic the 28 separate images together.

The resulting image has an rms which generally lies in the range 20-60 μJy across the field, with a spatial resolution of about 11 x 5 arcsec. A representative region of the image is shown in the left-hand panel of Fig. 2.

To illustrate the advantage of super-uniform weighting, we also show in Fig. 2 the same image made with natural weighting. The naturally-weighted image has a higher sensitivity but a poorer resolution, resulting in a significantly enhanced confusion rate when cross-identifying to the Spitzer observations.

## 2.2 Component Extraction

Most source extraction techniques, and all references to a "five-sigma" detection, implicitly assume a Gaussian noise distribution, which is unlikely to be found in radio interferometry images. Nevertheless, an assumption of Gaussian noise can simplify the initial stages of source extraction, and Bondi et al. (2003) have shown SExtractor2 (Bertin & Arnouts 1996) to generate a reliable noise image from radio data. We therefore used SEXtractor2 to produce a noise image, which we then divided into the mosaiced image to obtain a signal-to-noise ratio (S/N) image. The MIRIAD task *imsad* was then used to derive a preliminary list of component ''islands'' above a cut-off of 4 times the local rms noise. Each component island found by *imsad* was examined and refitted in the mosaiced image data with an elliptical Gaussian to derive component flux densities and sizes. All component images and fit parameters were inspected to check for obvious failures and poor fits that needed further analysis.

In some cases the automated process fitted a single Gaussian to a complex of two or more individual components, and so we re-fitted the data using several Gaussian components. In these cases, each component is listed as a separate component in subsequent analysis.

We then inspected the image of each resulting component. All components with either a peak flux or an integrated flux of less than 5 times the rms noise (as measured by Sextractor2) were discarded unless their morphology strongly suggested their reality, such as the extended source associated with the spiral galaxy S226 (see Section 3.6). Some parts of the ATLAS field contain artifacts which result in strongly non-Gaussian noise statistics, so that our simple noise cut-off fails to remove artifacts. Therefore, a subjective approach was used to remove any components which may have been generated or strongly affected by artifacts. The resulting sample therefore does not have a clearly-defined sensitivity limit, and cannot be used as a statistically complete sample.

In the region of the field where this catalogue overlaps with that of KAMNC there is, as expected, overall agreement, but also some significant differences. This is because of the super-uniform weighting scheme used here, as compared with the robust weighting scheme used by KAMNC, so the results presented here have lower sensitivity but higher resolution (and also, of course, a lower peak flux for extended sources). Other differences are attributable to different selection procedures. For example, ATCDFS_J033159.86-274541.3 occurs in the KAMNC catalog but not ours, because it lies on a grating ring (part of the pattern caused by calibration errors) of a nearby strong source and so was rejected by the artifact-removal procedure described above. Close inspection of the KAMNC data shows this component to be real, but it is still excluded from our catalog to maintain consistency.

Comparing the KAMNC fluxes with those presented here reveals the hazards of measuring fluxes in radio-astronomical images with different weighting and gridding schemes. While there is overall agreement, individual sources can differ significantly. For example, ATCDFS__J033219.82-274121.2 appears in the KAMNC catalogue with a flux density of 0.228 mJy, but was missed by our source extraction procedure. Examination of the image used for our source extraction shows this source to be visible, but with an integrated flux density of only 0.09 mJy, and so it was rejected by our source selection criterion, whereas it fell above the threshold in the more sensitive KAMNC observations. The difference between the two measured fluxes is consistent with the measured noise levels of the two images at this point (KAMNC place an error estimate of 0.085 mJy on this source).

Comparison of flux densities of all sources common to the two papers shows that our derived integrated flux densities tend to be 14% lower than KAMNC's. This may be partly attributable to the different weighting scheme, but there may also be a small overall calibration difference between the two sets of results. 27 of our sources also appear in the NRAO VLA Sky Survey (NVSS – Condon et al. 1988). The flux densities we measure are, on average, 11% lower than those listed in NVSS. However, the two surveys have very different synthesized beam sizes (NVSS 45 arcsec compared to ATLAS 6 arcsec) and so a simple comparison is unreliable.

These differences are, however, higher than expected. Some initial experiments to explore the source of this uncertainty have shown variations of about 10% in flux density depending on imaging parameters such as pixel size and weighting. We plan a set of extensive simulations to explore and understand these differences before the final data release of the ATLAS survey. In the meantime we assign a conservative estimated uncertainty of 20% to all flux densities, in quadrature with the rms noise of the image which is typically 40μJy. Formal flux density uncertainties from the fitting process are generally low compared to this uncertainty due to calibration, and so are not listed individually. We expect to derive more rigorous flux density uncertainty estimates in the final ATLAS catalogue.

### 2.3 Spitzer Observations

The Spitzer Space Telescope was launched in August 2003. It is equipped with several instruments, and here we use data taken with the InfraRed Array Camera (IRAC) at each of four bands (3.6, 4.5, 5.8, and 8 μm) and with the Multiband Imaging Photometer for Spitzer (MIPS) in its 24 μm band.

Before launch, proposals were invited for Legacy Science Programs, which were large coherent programs whose data would be of lasting importance to the broad astronomical community. One of the six Legacy projects chosen was the SWIRE (Spitzer Wide Extragalactic) Survey (Lonsdale et al 2003), which has observed a 6-square degree region surrounding the CDFS. The analysis of those data is described by Surace et al (2006), and have resulted in images to a depth of 5, 9, 43, 40, 193 μJy respectively in the four IRAC bands and in the 24μm MIPS band, together with a catalog of detected sources. Most of the SWIRE data are now in the public domain (see http://swire.ipac.caltech.edu/swire/astronomers/data_access.html). Here we use the SWIRE Public Data Release 3.

The IRAC and MIPS fluxes used here are, in the case of unresolved or noisy sources, aperture-corrected fluxes, as described by Surace et al. 2006. In the case of extended sources, Kron fluxes are used (Kron 1980, ApJS, 43, 305).

In some cases, we give infrared fluxes, and SWIRE identifications, for sources that do not appear in the public-release SWIRE catalogue, because the public catalog sensitivity is set at a more conservative threshold level than that used here. However, all SWIRE sources listed here are visible in the SWIRE images which are also in the public domain. Table 2 shows the approximate resolution and sensitivity at each wavelength resulting from the SWIRE survey.

## 2.4 Optical Photometry

In support of the SWIRE Legacy program, the CDFS field was observed for 18 nights with the MOSAIC II camera on the Blanco 4 m Telescope at Cerro Tololo. Fifteen pointings covered 4.5 $deg^2$ in four filters (U or u', g', r', i') to 5 sigma depths of ~24.5, 25.4, 25, 24 Vega magnitudes respectively, with an additional 1.5 $deg^2$ in z' to 23.3. Filter characteristics are given in Table 3. In addition, there is one deep pointing (25.2,25.7,25.5,24.5) covering 0.33 $deg^2$ centered at $03^h31^m14^s$ -28°$36^m$. The optical data overlap ~2.6 $deg^2$ of the primary radio data. Seeing ranged from 0.9-1.6 arcsec in mostly photometric weather. The pointings which were observed in photometric conditions were calibrated using photometric standards observed throughout the night. The non-photometric pointings were calibrated by cross-correlating sources which overlap regions of photometric fields. The calibration uncertainty is estimated to be 3% in g', r', i' and 5% in U, z'.

## 2.5 Spectroscopy

We obtained spectra of a subset of the radio sample in two separate observing sessions.

The first was part of the Australian Deep Radio Optical Infrared Target (ADROIT) survey, which used the 2dF multi-fiber spectrometer on the Anglo-Australian telescope (AAT) in the period 2003 November 19 to 25. The 316R (spectrograph 1) and 270R (spectrograph 2) gratings were used giving dispersions of 4.09 and 4.77A/pixel respectively. We split our targets into two samples, bright (R < 21) and faint (21 < R < 23).

The bright sample was observed in the normal 2dF observing mode with both spectrographs, while the faint sample was observed in nod-and-shuffle (N&S) mode (Glazebrook & Bland-Hawthorn, 2001). The bright field contained 64 targets going to spectrograph 1 and 65 going to spectrograph 2.

The faint configuration had 43 targets all of which went to spectrograph 1. The field was configured such that each object was allocated 2 fibers. Each pair of fibers was positioned at an A and B position, where A was the true position of the source and B was offset by 60 arcsec in Right Ascension (RA). Two of the four guide fibers were allocated to each of the A and B positions. A flat and arc were taken without the mask on and then the mask was positioned and the telescope slewed to the field. A second flat and arc were taken with the mask in place, and this was repeated every few hours between science observations. Each target exposure consisted of 60s at position A followed by 60s at position B repeated 15 times before the detector was read out. Each time the telescope was nodded between A and B, the charge would be shuffled back and forth in the detector by 50 pixels (in the Y direction). Each exposure therefore consisted of 900s at position A and 900s at position B. In total 19 N&S exposures were taken on the nights of the 19, 22, 23, 24 and 25 November. Some nights were affected by cloud and the seeing was variable, with a median of 2.1" (full range of 1.3 - 3.0). Therefore we obtained a total of 48600s (13.5 hours) exposure for each source (2x27x900s).

The data from the bright configuration were reduced using the standard *2dFdr* routines. The resulting mean signal-to-noise (S/N) was 16.4 (spectrograph 1) and 16.8 (spectrograph 2). The combined data were then run through a final program *NS_comb-AB* which combined the A and B observations of each target. The mean S/N in the final combined frames was 2.34 (*2dFdr* flux weighting) or 2.26 (*2dFdr* unweighted), from an effective exposure time of 34200s or 9.5 hours. The spectra were analyzed using the April 2004 version of the *runz* code used for the 2dFGRS and 2dF-SDSS LRG surveys.

The second set of spectroscopic observations was obtained using the AAOmega, the new spectrograph back-end to the AAT's 2dF multi-fiber spectrometer. AAOmega is a dual-arm bench-mounted spectrograph and provides greater throughput, stability and resolution than the previous 2dF spectrographs. Our observations were made on 2006 January 26-28, as part of the AAOmega Science Verification program.

We used the 580V and 385R volume-phase holographic gratings in the blue and red arms respectively providing a resolution of 1300 and dispersions of 0.1 nm/pixel (blue) and 0.16 nm/pixel (red). In both arms there are ~3.5 pixels per resolution element. Our observations were the first to make use of the nod-and-shuffle mode built into AAOmega. As above we allocated 2 fibres to each field, using a new version of the 2dF *configure* software applying a simulated annealing algorithm to maximize the number of fibre pairs configured (Miszalski et al. 2006). A total of 78 objects were configured. The A and B positions for the targets were separated by 120 arcsec and observations were carried out as described above, with each exposure consisting of 60s at position A followed by 60s at position B, then repeated 15 times. Thus each data frame has an effective exposure time of 1800s on target. Over the three nights we obtained 12 x 1800s on target in seeing of 1.2-1.5 arcsec, so a total of 6 hours on source integration was acquired. The median S/N ratios for objects in the R magnitude ranges 20.0-20.5, 20.5-21.0, 21.0-21.5, 21.5-22.0 and 22.0-22.5 are 10.6, 6.5, 3.1, 3.2 and 1.7 respectively. Redshifts were measured using the May 2006 version of the *runz* code, and their reliability assessed manually. 23 of the 25 objects at R<21 produced reliable redshifts, and 30 of the 47 targets at 21<R<23 produced reliable redshifts. One of the 6 objects fainter than R=23 had a reliable redshift.

# 3 Results and Analysis

## 3.1 The Image Data

Representative portions of the 20cm image of the CDFS-SWIRE field are given in Figs. 2 and 3. Postage stamp images of all sources will be available from the NASA extragalactic database on http://nedwww.ipac.caltech.edu/.

## 3.2 The Component Catalogue

Radio components were extracted from the image as described in Section 2.2. Table 4 shows the resulting catalog, which contains 784 radio components. The fields of Table 4 are as follows:

Column (1): Component Number: This is the internal designation of the component used within this paper
Column (2): Name: Designation for this radio component. In the case of single-component sources, this is identical to the source name used in table 5. This is the formal IAU designation and should be used in literature when referring to this component.
Column (3): Right ascension (J2000.0)
Column (4): Declination (J2000.0).
Columns (5) and (6): Rms uncertainties in Right Ascension and Declination. These include the formal uncertainties derived from the Gaussian fit together with a potential systematic error in the position of the calibrator source of 0.1 arcsec. Comparison of our positions with Spitzer positions in Section 3.3 below shows that these estimated uncertainties are realistic.
Column (7): Peak flux density at 20 cm (in mJy) of the fitted Gaussian component. The estimated uncertainty is 20% in quadrature with the rms given in column 12.
Column (8): Integrated flux density at 20 cm (in mJy) of the fitted Gaussian component. The estimated uncertainty is 20% in quadrature with the rms given in column 12.
Columns (9) and (10): Deconvolved FWHM major and minor axes (in arcsec) of the fitted Gaussian. If the undeconvolved fitted major or minor axis size was within one formal standard error of the restoring beam size, it was set to zero.
Column (11): Major-axis position angle (in degrees east of north)
Column (12): Rms: The value (in mJy) of the rms map generated by Sextractor at the position of the component.
Column (13): Comment

### *3.3 Radio-infrared cross-identifications, and identification of multi-component sources*

Although the spatial resolution of the radio image is typically $\theta_{HPBW} \sim 6$ arcsec, the positional error for an unresolved source in the presence of Gaussian noise is expected to be of the order of $\theta_{HPBW}$/SNR (Condon, 1997), where SNR is the signal-to-noise ratio, and should therefore be of the order of one arcsec or less for the unresolved radio components discussed in this paper. This expectation is largely confirmed by the source statistics discussed below. The SWIRE positional errors were in all cases less than one arcsec.

We note that other authors (e.g. de Ruiter et al 1977) have used automated techniques for performing cross-identifications and estimating the error rate, but we do not consider these techniques to be appropriate here because of the presence of sidelobes, extended sources, and multiple components, which make automated techniques less reliable. In most cases discussed here, the identification is unambiguous because of the relatively low density (compared to the source size and positional accuracy) of sources in both the radio and SWIRE images, and we estimate the error rate by repeating the identification process using spatially-shifted data.

For the cross-identifications, we used a pre-release version of the SWIRE Public Data Release 3 Catalog. The version that was publicly released is slightly more conservative than this pre-release version, and so some identifications do not appear in the public data release. However, all are visible on the publicly-released SWIRE images.

For each radio component, we initially searched the SWIRE catalogue for a source at any Spitzer wavelength within 3 arcsec, and counted the nearest such source as a correct identification, unless subsequently re-classified on the basis of morphology (see below). This resulted in a distribution of distance to the nearest identification shown in Table 5. About half the radio components have a SWIRE source within 1 arcsec, and 79% have a SWIRE source within 3 arcsec. For the 163 radio components that did not have an identification within 3 arcsec, we examined each source in turn, using both the radio and infrared images.

In 46 cases, the radio components formed a classical double radio source, in which the host galaxy lies roughly midway between the two radio lobes. A further 33 components were members of classical radio triple sources, in which a SWIRE source was coincident with the centre radio component, and no SWIRE source was visible coincident with the lobes. In 31 cases, the radio component appeared to be associated with a SWIRE source even though their central positions differed by more than 3 arcsec, typically because the source was extended. In 17 cases, a source was visible on the Spitzer images but had not been included in the SWIRE catalog either because it was too faint or because it was confused by a nearby bright source, and in five cases the radio source lay in a region which was outside the area observed in the SWIRE project. These statistics are summarized in Table 5.

As a further check on multiple sources, quite independently of the above process, we applied the technique described by Magliocchetti et al. (1998), in which a pair of radio components are classified as the lobes of a double radio source if they satisfy the criteria (a) $\theta < 100\sqrt{(S/100)}$, where S is their combined total flux in mJy and $\theta$ is their separation in arcsec, and (b) $0.25<S_1/S_2<4$, where $S_1$ and $S_2$ are the integrated flux densities of the two components. This technique has the advantages of being objective, and of having been demonstrated to work well at mJy flux densities. It has the disadvantages that it is untested at the flux densities observed here (where interacting starburst galaxies, for example, are far more common than at higher flux density levels) and that it does not make use of the additional information available from the Spitzer data. This is discussed further in Section 3.6 below.

Nevertheless, the two techniques show remarkable agreement. The Magliocchetti et al. test (hereafter called the M-test) missed only 5 of the 34 sources classified as triples or doubles by the subjective technique, and found a further 2 doubles which had been missed by the subjective technique. In addition, it identified 13 groups of sources which had been recognized as potential double or multiple sources in the subjective technique, but which had subsequently been classified as neighbors, clusters, or interacting galaxies, on the basis of their morphology or Spitzer identifications. For example, the pair of components C072 and C073 survives the M-test, but both have sub-arcsec Spitzer identifications. The probability of two sources both having spurious sub-arcsec Spitzer identifications is $4.10^{-4}$, and so we classify these as two separate galaxies. All cases where the two techniques disagreed were re-examined, and the verdict noted in the comments column of Table 6.

At the end of this cross-identification process, 22 radio sources remained which did not have SWIRE counterparts. While the weakest of these might be due to noise peaks in the radio image which had falsely been counted as radio sources, there remain eight sources where there is an unambiguous radio source with an integrated flux of at least 1 mJy, and where a good SWIRE image at that position shows no indication of an infrared source. We call these sources "Infrared-faint radio-sources" (IFRS), and will discuss them in more detail in Section 4.3 below.

We estimated the probability of false cross-identifications by shifting all the radio sources by one arcmin and repeating the process. As a result, we estimate that 15.25 of the 393 cross-ids within 1 arcsec (i.e. 3.9%) are false, 19.20 of the 167 (11.5%) between 1-2 arcsec are false, and 21.20 of the 57 (37.2%) between 2-3 arcsec are false.

After grouping together multiple radio components that appear to be physically part of one radio source as described above, the 784 radio components correspond to 726 radio sources (including the IFRS), which are identified with 682 SWIRE sources. We estimate 56 of these (or 8.2%) to be spurious identifications caused by confusion in the IRAC data.

Any systematic error in our positions should show up as a systematic offset between our positions and the Spitzer positions. The mean position offset between Spitzer and radio positions, averaged over all sources with a Spitzer identification, is 0.06 and 0.11 arcsec in Right Ascension and Declination respectively. Restricting the comparison to sources where the Spitzer identification is within 3 arcsec of the radio source, gives mean offsets of 0.04 and 0.09 arcsec respectively.

After the cross-identification process was completed, the SWIRE public data release 3 became available, which is slightly more conservative than the version used for the cross-identification, and does not include some faint or confused sources. Of the 682 identifications with previously catalogued sources, 82 did not appear in the SWIRE 3 public data release. In each of these cases, the SWIRE images were re-examined, and in 74 cases the identification with the previously catalogued source was confirmed. These sources do not have a formal "SWIRE3_J…" designation, and instead appear in Table 6 with their internal identification numbers in brackets. In the re-appraisal of the remaining eight cases, the identification was changed to the SWIRE source that appears in the public data release.

### *3.4 Source characterization*

Sources were characterized using the following criteria:

a. If the radio morphology indicates a classical radio double or triple source, or a core-jet source, then it is classified as an AGN. 35 sources were classified in this way.
b. If the radio source is stronger than the radio-FIR correlation by a factor of 10 (i.e. $\log (S_{24\mu m}/S_{20cm}) < 0$), then it is classified as an AGN. 113 sources were classified in this way. We note that all sources classified as AGN by criterion (a) which had a measured 24μm flux would also have been classified as an AGN by this criterion. Despite potential K-corrections to both the radio and infrared fluxes, the slope of the radio-FIR correlation does not appear to vary strongly with redshift (Appleton et al. 2004; Higdon et al. 2005), and so this criterion should be useful at all redshifts.
c. If a source is classified by other authors (either Afonso et al. (2006) or Croom et al. (2001)) on the basis of its spectroscopic or X-ray properties, then we assign it their classification. 8 sources were classified as AGN in this way, and 5 as star-forming galaxies.
s. Although the spectroscopy was mainly targeted at obtaining redshifts, some spectral classifications were obtained. Six sources were classified as AGN in this way, and two as star-formation galaxies
x. If a source was detected by Chandra (Giacconi et al. 2002) then in a few cases we can use its hardness ratio (HR) to classify it (Rosati et al 2002). All galaxies studied by Rosati et al. with HR>0.2 are classified as Type II AGN, so here we classify all such galaxies as AGN. Sources with HR<-0.2 are usually either a Type I AGN or a star forming galaxy, although a few Type II AGN also have this hardness ratio. Therefore, in cases where the optical image precludes a Type I AGN, we conclude that most such sources are likely to be star-forming galaxies, but this is far from an unambiguous classification. 3 sources were classified as AGN and 2 were confirmed as star-forming galaxies on the basis of their X-ray properties.

In addition, we note that sources could arguably be classified as AGN if their radio luminosity > $10^{24}$ $\mathrm{WHz}^{-1}$, as all well-known galaxies at this radio luminosity are AGN, but we have not done this as it pre-empts the possibility raised by Chapman et al. (2003a) and others that there may be super-starbursts with very high radio luminosities.

We also note that the above classification process can successfully identify AGNs, but is very inefficient at classifying star-forming galaxies. Thus, nearly all our classifications are of AGN.

However, Norris et al (in preparation) have shown that a significant fraction of the remaining galaxies have a spectral energy distribution (SED) characteristic of star formation. This confirms the findings of previous radio surveys which find both AGN and star-formation-driven objects represented in such samples. However, we emphasize that these classifications are heavily biased in favor of AGNs, and should not be used as an estimator of SF/AGN activity.

## 3.5 The Source Catalog

The Radio Source Catalog is presented in Table 6, with the following columns.

Column (1): Source Number: This is the internal designation of the source used within this paper

Column (2): Name: Designation for this radio source. In the case of single-component sources, this is identical to the component name used in Table 4. This is the formal IAU designation and should be used in literature when referring to this source.

Column (3): Component number(s) corresponding to Table 4.

Column (4): SWIRE ID Designated name of the SWIRE identification used in SWIRE Public Data Release 3. In cases where the source does not appear in the Public Data Release 3, but did appear in the pre-release catalogue, the source identification from the pre-release catalogue is shown in brackets. A blank indicates there is no catalogued SWIRE source, but a source may still be present in the SWIRE image, in which case it will be noted in the "Comments" column.

Columns (5) and (6): Right ascension (J2000.0) and Declination (J2000.0). In the case of a single component, this is the position of the radio source. In the case of a complex source, such as a radio double, this is the position of the host galaxy. In the latter case, this is the optical position if one is available, or else an infrared position.

Column (7): Total flux density at 20 cm (in mJy). This is the total integrated 20 cm flux of all components included in the source.

Columns (8) to (12): Infrared fluxes measured at 3.6μm, 4.5μm, 5.8μm, 8μm, 24μm, in μJy. These fluxes are optimized so that they are aperture extractions for point sources and extended (Kron) extractions for extended sources. -1 indicates that the source was undetected, and a blank indicates that the source was not observed, or that its flux is not listed in the SWIRE catalogue.

Columns (13) to (17): SDSS U/u′, g′, r′, I′, z′ are aperture magnitudes for stellar sources, and integrated magnitudes for extended sources. All are in the Vega system. Filter characteristics are shown in Table 3. 99 indicates that the source was undetected, and a blank indicates that the source was not observed.

Column (18): Spectroscopic redshift. In most cases these have been measured by us as part of this program, as described in Section 2.5. In some cases they are taken from other authors, in which case this is noted in the "comments" column. To avoid ambiguity, photometric redshifts are not included here

Column (19): ID type. The type and accuracy of the radio-infrared identification, using the code listed in table 5.

Column (20): Classification (AG/SF) based on the criteria described in Section 3.4.

Column (21): Basis for the classification. The criterion used for the classification is given by the lower-case letter: a=morphology (i.e. double, triple, or core-jet radio source), b=value of $q_{24\mu m} = \log(S_{24\mu m}/S_{20cm})$, c=classification taken from the literature, s=based on spectroscopy presented in this paper, x=based on X-ray hardness ratio given by Giacconi et al. (2002.)

Column (22): Comments. "M-test" refers to the criterion for selecting double radio sources described by Magliocchetti et al. (1998). "XIDnnn(mm)" indicates that the source was detected by Chandra (Giacconi et al 2002), and is labeled XIDnnn, with hardness ratio mm, in their catalogue. z(x) gives the reference for the redshift as follows: a=this paper (ADROIT observations), b= this paper (AAOmega observations), c=Afonso et al. (2006), d=Croom et al. (2001), e=da Costa et al. (1998), f=Vanzella et al. (2006), g=Colless et al.

(2001),h=Le Fevre et al. (2004), i=Loveday et al. (1996), j=Cimatti et al. (2004), k=Way et al. (2005), l= Lauberts & Valentijn (1989).

## *3.6 A representative sample of sources*

Here we present images of a small but representative sample of sources to illustrate the quality of the data, and also to illustrate the issues which impact on source identification. In each case, an image is shown in Fig. 3 which shows the 20 cm radio contours (with the lowest contour generally set at 100 µJy) overlaid on the 3.6µm SWIRE images. References to the spectral energy distribution (SED) are all taken from Norris et al. (in preparation).

### The S323 region

Fig. 3 (top) shows a region in which several types of object are visible. S323 = ATCDFS_J033117.00-275515.3 (C342, C346, C348) is a classical triple radio-galaxy, with a bright SWIRE galaxy coincident with its core. The M-test successfully identifies these three sources as associated. The morphology of this source (two bright radio lobes surrounding a bright SWIRE source) is unmistakable. Spectroscopy shows the host galaxy to lie at a redshift of 1.37, with the spectrum of a broad-line quasar.

To the North lies a single strong radio component (S331/C355 = ATCDFS_J033124.89-275208.3) which is coincident with a reasonably bright SWIRE source, which is also visible at optical wavelengths. This source has
$q_{24\mu m} = \log (S_{24\mu m}/S_{20cm}) = -2.0$, indicating that within the host galaxy lies an AGN.

S291/C311 = ATCDFS_J033055.63-275201.7 is an extended radio source, which may be a core/jet signifying an AGN. At its centre is a SWIRE source with a measured spectroscopic redshift of 0.3382, and the SED of an elliptical galaxy.

S279/C298 = ATCDFS_J033046.26-275517.5 is a single radio component coincident with a SWIRE source with the SED of a star-forming galaxy. The value of $q_{24\mu m}$ for this galaxy is also consistent with the radio-FIR correlation, supporting its identification as a star-forming galaxy.

S287 = ATCDFS_J033056.45-275508.0 is a linear arrangement of three radio components (C305, C307, C312), which are remarkably symmetrical both in spacing and in flux density. These three components fall just below the M-test criterion because their flux densities are relatively low for the measured separation between them. The centre component is coincident with a SWIRE galaxy with the SED of a star-forming galaxy, and a measured spectroscopic redshift of 0.8934, while the two outer components have no SWIRE identification, suggesting that this is a triple radio source. If this identification is correct, it suggests that an AGN is buried within a star-forming galaxy.

### The S226 galaxy

C244/S226 = ATCDFS_J032956.56-284632.6 is a diffuse 5 mJy radio source which is coincident with a bright barred spiral galaxy (ESO418-G007) at z = 0.037, shown in Fig. 3 (centre). 30 arcsec away is a compact (but slightly resolved) 4 mJy radio source (C241/S223) which is coincident with a bright compact infrared object, which appears to lie at the end of one of the spiral arms of S226.

We note that the M-test incorrectly classifies these two sources as a radio double.

### The S440 region

Fig 3 (bottom) shows the region around this source, and helps illustrate the nomenclature and conventions used in this paper. It also illustrates the shortcoming of the M-test when applied to deep surveys such as this.

Two nearby radio sources, S442 = ATCDFS_J033229.84-274423.8 and S443 = ATCDFS_J033229.97-274405.4, are clearly identified with bright SWIRE galaxies. These two sources were also observed by Afonso et al. (2006) who described them both as “flocculent” star-forming galaxies and measured a redshift for each of 0.076. They were also detected by Chandra (Giacconi et al 2002) as XID 95 and XID116, with hardness ratios of -0.7 and -0.l56 respectively, which, since their optical appearance precludes Type 1 AGN, suggests that they are star-forming galaxies (Rosati et al. 2002).

In the 3.6μm IRAC image in Fig 4b there appears to be a weak bridge of emission connecting them, which was also noted by Giacconi et al (2001), who classified them as “interacting”.
The M-test classifies these two galaxies as a radio double, which is clearly incorrect.

Two other radio sources, identified here as components C473 and C476, were also observed by Afonso et al. who were unable to identify them at optical wavelengths. Here, we also fail to find an identification at any of the SWIRE wavelengths. However, we note that they appear to be connected by a bridge of radio emission, and that there is a bright SWIRE source (SWIRE3_J033228.79-274356.1) between them, and within the radio contours, and so we tentatively identify this as the host galaxy of a double radio source, and so we designate the two components C473 and C476 as one radio source S440 = ATCDFS_J033228.79-274356.1. The M-test correctly classifies these two galaxies as a radio double. The central host galaxy is also detected by Chandra (XID103) with HR=-0.69, suggesting this may be a Type I AGN.

In addition, the M-test classifies C473 and C477, and C476 and C478, respectively, as radio doubles. It is clear that, although the test has been demonstrated to work well at high flux densities, it is less successful at the flux densities observed here, because galaxies are far more likely to have nearby companions than at higher flux density levels. The reliability of classification is greatly increased by referring to data at other wavelengths, such as the Spitzer data.

# 4 Discussion

## 4.1 Starburst or AGN?

It is well established (e.g. Dickey & Saltpeter 1984, de Jong et al 1985) that in nearby galaxies dominated by star formation, the radio and far-infrared (FIR) emission are strongly correlated., and Appleton et al. (2004) have shown that the correlation is still valid at high redshifts. Appleton et al. also showed that the correlation can be seen, albeit with a higher scatter, in plots of 24μm (as opposed to FIR) flux against radio flux. Boyle et al. (2006) have shown that the correlation is also present in stacked radio images down to microJansky levels.

Luminous radio galaxies and AGN depart very strongly from the correlation (e.g. Sopp & Alexander, 1991), making the correlation a good test for AGN. A source which departs from the correlation is likely to be an AGN, but it cannot be concluded that a source which follows the correlation is not an AGN, because Roy et al. (1998) showed that most Seyfert galaxies also followed this correlation, suggesting that, despite the presence of an AGN, their radio luminosity is still dominated by star formation activity.

In Fig. 4 we plot the observed integrated 20 cm radio flux against the observed 24 μm infrared flux for all our identified sources which have measured 24μm fluxes. No k-correction has been performed here, or elsewhere in this paper, because redshifts and spectral shapes are generally too poorly known for this sample to do so with confidence. It is clear that all sources that have been classified as AGN on the basis of their morphology depart very strongly from the correlation.

In the lower right of the diagram is a paucity of sources, and the sharp diagonal boundary between this space and the plotted points is close to the radio-FIR correlation. Our sample of sources clearly fails to follow the radio-FIR correlation, presumably because of the large numbers of radio-luminous AGN in the sample. However, Fig. 4 shows that the correlation is close to a lower bound to the value of $S_{20cm}/S_{24\mu m}$.

This is more clearly demonstrated in Figure 5, which shows a histogram of $q_{24\mu m} = \log S_{24\mu m}/S_{20cm}$ for our sample and those of Higdon et al. (2005), which is dominated by obscured AGN, and Appleton et al. (2004), which is dominated by star-forming galaxies (because their sample had been color-selected to maximize the number of star-formation galaxies at around z=1).

The sharp cut-off on the right is caused by the significant absence of sources below the line in Figure 5, and is common to all such radio surveys. The tail (and possibly the peak) to the left of the plot indicate sources which are radio-bright, and presumably driven by AGN. Our sample clearly overlaps both those of Appleton et al. and Higdon et al.. Presumably those overlapping the Appleton et al. curve are primarily powered primarily by star formation activity, while those overlapping the Higdon curve are powered primarily by AGN activity. It should be noted that less than half of our radio sources are detected at 24μm in the SWIRE survey (although nearly all are detected at shorter wavelengths), and so there is a large undetected population of sources to the left of this plot.

Thus, although no quantitative conclusions can be drawn from this study, it is clear that our sample contains significant numbers of both AGN and star-forming galaxies. This is supported by the X-ray and other indicators referred to in Table 6.

### *4.2 Redshift distribution*

In Figure 6 we show the distribution of spectroscopic redshifts for objects in our sample. Only 7 of the sources in our sample are classified as star formation galaxies, which is too small to be usefully shown in this histogram, but we expect that most of the objects which have not been classified as AGN are powered primarily by star formation.

The highest redshift object in our sample is at z=2.18, and all objects at redshifts > 1.2 are classified as AGN. However, this is strongly influenced by selection effects, because only the optically brighter (typically R < 22.5) galaxies have so far had their redshifts measured, and AGNs tend to have more prominent emission lines. Photometric redshifts for this sample (Norris et al, in preparation) extend to significantly higher redshifts for both AGN and non-AGN galaxies.

Nevertheless it is clear that the number of detected galaxies declines with redshift, as expected, with the exception of a pronounced maximum in both AGN and non-AGN galaxies at about z=0.7, presumably due to large-scale structure, such as the clusters at z=0.66 (Croom et al., 2001), and z=0.73 (Gilli et al., 2003). We also note the flat tail extending to high redshifts, which consists entirely of AGN.

### *4.3 Infrared-Faint Radio Sources (IFRS)*

Richards et al (1999) found that 20% of the microJansky radio sources in the HDF-N had no counterpart brighter than I=25. Further observations showed that several of these were very red, with I-K>4. Norris et al. (2005) and Huynh et al. (2005) found a similar result in the HDF-S. For example, the strongest radio source in the HDF-S (ATHDFS_J223258.5-603346) is extremely faint (V=27.05) and red (I-K=3.45), and is also unusually radio-loud (log(S20/I)=3.74). However, whether the radio emission is being produced by star formation or by an AGN, we expect that the dust that is apparently hiding the radio-producing activity should be bright at mid-infrared wavelengths. Thus we expected that all radio sources detected by ATLAS would appear in the SWIRE catalog.

Unexpectedly, we find that a small number of radio sources in our sample are not visible at any Spitzer wavelength. We denote this rare class of objects "Infrared-Faint Radio Sources" (IFRS).

There are 22 such objects in our sample. While the weakest of these may be ascribed to statistically unusual noise peaks or imaging artifacts (although we have attempted to remove all such spurious sources), some of them are as strong as 5 mJy, and their reality is beyond question. Figure 7 shows two examples. In both cases, the sources are invisible in all Spitzer infrared wavebands (optical identification is limited by confusion) and so the only information on these sources comes from the radio.

Fig. 8 shows a "stacked" IRAC image for all the 22 IFRS, which has been obtained by summing the individual 3.6 μm images centred on the 22 IFRS radio positions. We also show a stacked image of 3.6 μm images centred on the eight strongest IFRS, in case the weaker ones are radio artifacts. No source is detected at the radio position at any Spitzer waveband in either of the stacked images, implying that the mean flux of these is at least a factor of $\sqrt{22}$ and $\sqrt{8}$ respectively below the SWIRE sensitivity limit (Table 2). A similar result has been obtained at each of the other SWIRE bands. If the IFRS simply represented a tail to the observed distribution of radio/infrared flux densities, then they would be expected to fall just below the SWIRE sensitivity limit, and should appear in the stacked image. Their absence from the stacked image suggests either that the distribution of radio/infrared flux density ratios is bimodal or else that it has a tail extending to high values of that ratio.

Possible explanations for these sources include:

- An AGN or SF galaxy so heavily obscured, or at such a high redshift, that all its dust emission is radiated at far-infrared wavelengths beyond 24μm, and thus undetectable by Spitzer. This model must also accommodate one case (S415) of a Chandra non-detection.
- A starburst or AGN in a transitory phase in which electrons are producing radio emission, but there is insufficient dust to produce detectable infrared emission.
- A radio lobe from an unidentified radio source.
- Some other exotic object, which may be galactic.

Higdon et al. (2005) have identified a related class of sources which they denote "Optically Invisible Radio Sources" (OIRS). The OIRS they identify are 20cm radio sources observed with the Very large Array (VLA) in the Bootes field which do not have an optical identification at B, R, or I bands. Most of the OIRS also do not have a 24μm detection at a sensitivity level of 0.3 mJy, which is similar to the SWIRE 24μm sensitivity.

Assuming the invisibility of both IFRS and OIRS is caused by dust extinction, the IFRS have a more extreme selection criterion than OIRS, in that we require no detection at any of the Spitzer bands, rather than at the shorter wavelengths required by OIRS. Although most of the OIRS sources do not have 24μm counterparts, the shorter SWIRE bands are generally more sensitive both to AGN and star formation galaxies, and approximately half the radio sources presented in this paper do not have 24μm counterparts. Thus we expect that the IORS and IFRS overlap, with the IFRS generally being more extreme examples of OIRS.

IFRS will be discussed at greater length by Norris et al. (in preparation).

# 5 Conclusion

We have presented data for a sample of about 800 radio components in the CDFS field, even though we are only about half-way through our radio survey of this region, primarily to facilitate further work at other wavelengths. Because this sample is not yet statistically complete, we have restricted the discussion on the astrophysical implications. Nevertheless, we may draw some preliminary conclusions.

Some of the galaxies have radio data which show an unmistakable signature of an AGN, either because their radio/24 μm ratio departs from the expected correlation, or because their radio morphology indicates classical radio doubles or triples. While some of these have also been identified as AGN on the basis of optical or X-ray data, many have not, demonstrating the value of radio observations as a technique for identifying AGN. In particular, some galaxies which we have classified as AGN have not been detected by Chandra, and we note that other authors (Alonso-Herrero et al 2005, Donley et al 2005, Rigby et al 2005) have also reported radio-selected AGNs which have not been detected by Chandra.

About half the ATLAS radio sources lie close to the radio-FIR correlation, and are presumably driven primarily by star-formation activity. Thus the ATLAS radio sources include comparable numbers of both star-forming galaxies and AGNs. However, all galaxies with a measured redshift greater than 1.2 have been classified as AGN, from which we conclude that, at the current level of sensitivity, star-forming galaxies are mainly confined to redshifts less than about 1. We find no evidence for high-luminosity star-forming galaxies at $z>1$, although we acknowledge that this may be partly attributed to the difficulty of measuring redshifts of galaxies other than AGN at high redshifts.

We have also identified a class of radio sources, the infrared-faint radio sources, which are invisible at optical and infrared wavelengths. These objects are rare (a few per square degree at current sensitivity levels) and so can only be found in wide deep surveys such as this. We expect to find more of these objects as we continue to increase the sensitivity of the ATLAS survey by adding the observations that are planned over the next 1-2 years.

## *Acknowledgements*

We thank Jim Condon for helpful comments on this paper. RPN gratefully acknowledges “Visiting Scientist” support from the Spitzer Science Center, JA gratefully acknowledges the support from the Science and Technology Foundation (FCT, Portugal) through the research grant POCTI/CTE-AST/58027/2004, and IRS acknowledges support from the Royal Society. The Australia Telescope Compact Array is part of the Australia Telescope which is funded by the Commonwealth of Australia for operation as a National Facility managed by CSIRO. This research has made use of the NASA/IPAC Extragalactic Database (NED) which is operated by the Jet Propulsion Laboratory, California Institute of Technology, under contract with the National Aeronautics and Space Administration.

# Figures and Tables

### *Table 1: Summary of radio observations*

| Date | Project ID | Configuration | Hours on source |
|---|---|---|---|
| 4-27 Apr 2002 | C1035 | 6A | 99 |
| 24-29 Aug 2002 | C1035 | 6C | 50 |
| 7-12 Jan 2004 | C1241 | 6A | 26 |
| 1-5 Feb 2004 | C1241 | 6B | 27 |
| 6-12 Jun 2004 | C1241 | 750D | 63 |
| 24-30 Nov 2004 | C1241 | 6D | 57 |
| **TOTAL** | | | **322** |

### *Table 2: SWIRE resolution and sensitivity.*

| Band | Limiting (5σ) sensitivity (μJy) | FWHM Resolution (arcsec) |
|---|---|---|
| 3.6 μm | 5 | 1.2 |
| 4.5 μm | 9 | 1.2 |
| 5.8 μm | 43 | 1.2 |
| 8 μm | 40 | 2.0 |
| 24 μm | 192.5 | 5.5 |

### *Table 3: Filter Characteristics used for Optical Photometry*

| Filter | NOAO Name | Effective Wavelength (nm) | Limiting magnitude | Vega to AB Conversion |
|---|---|---|---|---|
| u' | | 361.8 | 24.5 | 0.90 |
| U | c6001 | 366.7 | 24.5 | 0.73 |
| g' | c6017 | 476.4 | 25.4 | -0.10 |
| r' | c6018 | 627.9 | 25 | 0.16 |
| i' | c6019 | 764.7 | 24 | 0.39 |
| z' | c6020 | 869.0 | 23.3 | 0.55 |

Notes:
(1) All magnitudes given in this paper are in the Vega system. They may be converted to AB magnitudes by adding the number in the last column.
(2) The limiting magnitudes are the 5-sigma limits obtained over most of the field. Deeper limits were obtained in the region of the GOODS field, as described in Section 2.4.

## *Table 4: Catalog of radio components*

The full table will be published in electronic form only.
Here we show a sample.

Table 4: Component catalogue

| # | Name | Radio RA hhmmss | Radio dec ddmmss | err(RA) arcsec | err(dec) arcsec | Peak Fl mJy | Int flux mJy | Bmaj arcsec | Bmin arcsec | Bpa deg | rms (uJy/beam) | Comment |
|---|---|---|---|---|---|---|---|---|---|---|---|---|
| C001 | ATCDFS_J032602.78-284709.0 | 3:26:02.785 | -28:47:09.068 | 0.78 | 0.73 | 0.70 | 1.38 | 8.3 | 2.6 | 60.8 | 79 | |
| C002 | ATCDFS_J032604.15-275659.3 | 3:26:04.152 | -27:56:59.392 | 0.55 | 0.90 | 0.71 | 1.97 | 11.5 | 7.3 | -17.0 | 72 | |
| C003 | ATCDFS_J032605.68-274734.4 | 3:26:05.685 | -27:47:34.485 | 0.03 | 0.05 | 40.81 | 74.70 | 5.9 | 5.6 | 85.7 | 119 | |
| C004 | ATCDFS_J032606.95-275332.2 | 3:26:06.955 | -27:53:32.260 | 0.51 | 1.19 | 0.41 | 0.43 | 0.0 | 0.0 | -1.0 | 77 | |
| C005 | ATCDFS_J032611.47-273243.8 | 3:26:11.475 | -27:32:43.817 | 0.02 | 0.03 | 69.65 | 110.90 | 5.3 | 3.3 | 89.6 | 156 | |
| C006 | ATCDFS_J032613.70-281717.7 | 3:26:13.701 | -28:17:17.715 | 0.56 | 0.78 | 0.48 | 0.54 | 0.0 | 0.0 | 0.0 | 78 | |
| C007 | ATCDFS_J032615.48-284629.2 | 3:26:15.489 | -28:46:29.245 | 0.32 | 0.35 | 0.45 | 0.71 | 0.0 | 0.0 | 0.0 | 66 | |
| C008 | ATCDFS_J032615.55-280601.0 | 3:26:15.557 | -28:06:01.050 | 0.28 | 0.46 | 0.73 | 1.06 | 0.0 | 0.0 | -1.0 | 57 | |
| C009 | ATCDFS_J032616.35-280014.6 | 3:26:16.353 | -28:00:14.618 | 0.15 | 0.28 | 1.24 | 1.66 | 0.0 | 0.0 | -1.0 | 60 | |
| C010 | ATCDFS_J032616.41-271621.1 | 3:26:16.419 | -27:16:21.106 | 0.12 | 0.21 | 4.17 | 7.84 | 9.1 | 2.8 | -40.0 | 90 | |
| C011 | ATCDFS_J032617.43-280709.9 | 3:26:17.430 | -28:07:09.955 | 0.16 | 0.22 | 6.73 | 13.83 | 8.6 | 3.3 | 58.2 | 58 | S lobe of radio double |
| C012 | ATCDFS_J032618.22-280703.5 | 3:26:18.225 | -28:07:03.541 | 0.46 | 0.45 | 3.80 | 10.29 | 11.2 | 3.9 | 70.0 | 57 | N lobe of radio double |
| C013 | ATCDFS_J032622.07-274324.4 | 3:26:22.079 | -27:43:24.483 | 0.03 | 0.06 | 17.35 | 27.81 | 5.5 | 4.4 | 35.4 | 72 | |
| C014 | ATCDFS_J032625.10-280908.8 | 3:26:25.109 | -28:09:08.856 | 0.62 | 0.85 | 0.41 | 0.96 | 9.0 | 5.0 | 62.1 | 57 | |
| C015 | ATCDFS_J032626.90-275610.9 | 3:26:26.904 | -27:56:10.916 | 0.09 | 0.18 | 2.65 | 4.14 | 6.6 | 3.6 | -26.1 | 63 | |

***Table 5: Cross-identifications and expected false cross-ID rate.***

| ID type | description | #cpts | #srcs | #spurious |
|---|---|---|---|---|
| 1 | within 1 arcsec of SWIRE source | 393 | 393 | 15.3 |
| 2 | 1-2 arcsec of SWIRE source | 167 | 167 | 19.2 |
| 3 | 2-3 arcsec of SWIRE source | 57 | 57 | 21.2 |
| 4 | good ID but >3 arcsec | 31 | 31 | 0 |
| -1 or 5 | part of a classical radio double | 46 | 23 | 0 |
| 6 | part of a classical radio triple | 33 | 11 | 0 |
| 7 | irac src not in SWIRE catalog | 17 | 17 | 0 |
| 8 | outside SWIRE region | 5 | 5 | 0 |
| 9 | IFRS | 22 | 23 | 0 |
| 10 | part of another source (e.g. knots in jets) | 13 | 0 | 0 |

## *Table 6: Catalog of Radio Sources with their identifications and classifications*

The full table will be published in electronic form only.  Here we show a sample.

**Table 6: Source catalogue**

| # | Name | Component number(s) | SWIRE ID | Radio RA hhmmss | Radio dec ddmmss | 20cm flux (mJy) | 3.6µm flux (µJy) | 4.5µm flux (µJy) | 5.8µm flux (µJy) | 8.0µm flux (µJy) | 24µm flux (µJy) | U | G | R | I | Z | z(sp) | ID type | Class | Basis | Comments |
|---|---|---|---|---|---|---|---|---|---|---|---|---|---|---|---|---|---|---|---|---|---|
| S130 | ATCDFS_J032822.70-283157.7 | C140 | SWIRE3_J032822.69-283157.9 | 3:28:22.701 | -28:31:57.774 | 1.3 | 3948.5 | 2722.9 | 3570.3 | 21307.8 | 15713.7 | 18.7 | 17.9 | 17.1 | 16.4 | | | 1 | | | |
| S131 | ATCDFS_J032823.93-281519.8 | C142 | SWIRE3_J032823.95-281520.0 | 3:28:23.931 | -28:15:19.848 | 0.6 | 822.8 | 888.4 | 1225.0 | 4898.1 | 11087.7 | | | | | | | 1 | | | |
| S132 | ATCDFS_J032824.45-281837.5 | C143 | SWIRE3_J032824.52-281839.8 | 3:28:24.454 | -28:18:37.599 | 0.2 | 165.9 | 232.6 | 340.5 | 489.2 | 2290.4 | | | | | | | 3 | | | |
| S133 | ATCDFS_J032824.56-284021.7 | C144 | SWIRE3_J032824.56-284021.6 | 3:28:24.567 | -28:40:21.796 | 11.1 | 14.7 | 13.2 | -1.0 | -1.0 | -1.0 | 99.0 | 99.0 | 99.0 | 99.0 | | | 1 | | | |
| S134 | ATCDFS_J032824.71-274149.8 | C145,C147 | SWIRE3_J032824.71-274149.3 | 3:28:24.712 | -27:41:49.855 | 1.4 | 46.4 | 55.0 | 71.4 | -1.0 | 505.2 | 99.0 | 99.0 | 99.0 | 99.0 | | | 1 | AGN | b | Core-jet |
| S135 | ATCDFS_J032825.37-274445.5 | C146 | SWIRE3_J032825.35-274445.0 | 3:28:25.373 | -27:44:45.502 | 1.0 | 30.4 | 32.7 | 43.8 | -1.0 | -1.0 | | | | | | | 1 | | | |
| S136 | ATCDFS_J032825.92-271701.3 | C141,C148,C151 | SWIRE3_J032825.92-271701.3 | 3:28:25.92 | -27:17:01.32 | 34.2 | 38.1 | 25.9 | -1.0 | -1.0 | -1.0 | | | | | | | -1 | AGN | a | Radio double with connecting jet |
| S137 | ATCDFS_J032826.50-281920.5 | C149 | SWIRE3_J032826.51-281920.7 | 3:28:26.504 | -28:19:20.578 | 1.2 | 589.1 | 483.3 | 146.7 | 133.9 | 580.0 | 99.0 | 21.9 | 20.2 | 99.0 | | 0.4265 | 1 | AGN | b | z(b) |
| S138 | ATCDFS_J032826.55-273304.2 | C150 | SWIRE3_J032826.52-273304.1 | 3:28:26.556 | -27:33:04.258 | 4.1 | 105.0 | 83.2 | 51.6 | 41.2 | -1.0 | 99.0 | 99.0 | 99.0 | 22.7 | | | 1 | | | |
| S139 | ATCDFS_J032829.30-280151.0 | C152 | SWIRE3_J032829.30-280150.5 | 3:28:29.309 | -28:01:51.019 | 1.6 | 294.4 | 166.4 | 106.5 | 72.1 | -1.0 | 99.0 | 24.5 | 22.3 | 20.8 | | 0.9010 | 1 | | | z(b) |
| S140 | ATCDFS_J032832.77-273538.7 | C153 | SWIRE3_J032832.78-273540.2 | 3:28:32.773 | -27:35:38.763 | 0.6 | 238.1 | 174.7 | 176.2 | 2183.9 | 3603.7 | 20.2 | 19.9 | 19.3 | 18.7 | | | 2 | | | |
| S141 | ATCDFS_J032832.79-285536.3 | C154 | SWIRE3_J032832.78-285536.1 | 3:28:32.797 | -28:55:36.347 | 17.3 | 42.9 | 49.5 | 49.5 | -1.0 | -1.0 | 99.0 | 24.3 | 23.6 | 22.5 | | | 1 | | | |
| S142 | ATCDFS_J032833.79-280152.5 | C155 | SWIRE3_J032833.78-280153.5 | 3:28:33.798 | -28:01:52.559 | 0.3 | 19.8 | 18.4 | -1.0 | -1.0 | -1.0 | 99.0 | 99.0 | 99.0 | 99.0 | | | 1 | | | |
| S143 | ATCDFS_J032835.63-273515.1 | C156 | SWIRE3_J032835.53-273514.2 | 3:28:35.631 | -27:35:15.142 | 3.2 | 36.1 | 36.0 | -1.0 | -1.0 | -1.0 | 99.0 | 99.0 | 99.0 | 99.0 | | | 2 | | | |
| S144 | ATCDFS_J032836.22-271650.3 | C157 | SWIRE3_J032836.19-271650.3 | 3:28:36.220 | -27:16:50.387 | 0.8 | 52.2 | 54.6 | -1.0 | 37.4 | -1.0 | | | | | | | 1 | | | |
| S145 | ATCDFS_J032836.53-284156.0 | C158,C159 | SWIRE3_J032836.52-284156.0 | 3:28:36.53 | -28:41:56.00 | 1357.8 | 4030.0 | 5687.8 | 9769.3 | 24294.5 | 202434.7 | 18.6 | 18.3 | 17.6 | 16.7 | | | -1 | AGN | ab | Radio double. Strongest radio source in field. Failed M-test because of flux ratio. |
| S146 | ATCDFS_J032837.04-275434.9 | C160 | SWIRE3_J032837.04-275435.0 | 3:28:37.049 | -27:54:34.939 | 0.6 | 776.0 | 513.2 | 731.6 | 4758.2 | 6575.6 | 18.6 | 18.5 | 18.0 | 17.5 | | | 1 | | | |
| S147 | ATCDFS_J032840.34-280539.9 | C161 | SWIRE3_J032840.34-280539.2 | 3:28:40.342 | -28:05:39.981 | 0.6 | 17.1 | 20.7 | -1.0 | -1.0 | 199.2 | 99.0 | 99.0 | 99.0 | 23.6 | | | 1 | AGN | b | |
| S148 | ATCDFS_J032841.10-283644.3 | C162 | SWIRE3_J032841.23-283646.6 | 3:28:41.102 | -28:36:44.369 | 2.2 | 113.2 | 56.2 | 52.9 | 34.3 | -1.0 | 99.0 | 24.5 | 22.9 | 21.5 | | | 4 | | | |
| S149 | ATCDFS_J032842.40-274447.7 | C163 | SWIRE3_J032842.41-274446.5 | 3:28:42.401 | -27:44:47.799 | 0.3 | 42.9 | 36.8 | 37.1 | -1.0 | -1.0 | | | | | | | 2 | | | |
| S150 | ATCDFS_J032843.38-282157.7 | C164 | SWIRE3_J032843.33-282157.2 | 3:28:43.386 | -28:21:57.758 | 1.3 | 203.0 | 154.4 | 98.9 | 48.2 | -1.0 | 99.0 | 22.0 | 20.3 | 19.2 | | 0.4294 | 1 | | | z(b) |
| S151 | ATCDFS_J032844.28-282323.0 | C165 | SWIRE3_J032844.22-282323.5 | 3:28:44.288 | -28:23:23.045 | 1.0 | 38.2 | 52.6 | 66.1 | 86.0 | 409.2 | 99.0 | 99.0 | 99.0 | 99.0 | | | 1 | AGN | b | |
| S152 | ATCDFS_J032846.65-282616.6 | C166 | SWIRE3_J032846.56-282618.1 | 3:28:46.651 | -28:26:16.643 | 58.0 | 96.4 | 84.5 | 102.2 | 83.9 | -1.0 | 99.0 | 99.0 | 99.0 | 99.0 | | | 2 | | | |
| S153 | ATCDFS_J032847.23-271512.9 | C167 | SWIRE3_J032847.21-271513.1 | 3:28:47.231 | -27:15:12.932 | 6.0 | 12.3 | 19.1 | -1.0 | -1.0 | -1.0 | | | | | | | 1 | | | |
| S154 | ATCDFS_J032848.75-283523.6 | C168,C169 | SWIRE3_J032848.75-283523.6 | 3:28:48.75 | -28:35:23.68 | 4.2 | 35.2 | 30.7 | -1.0 | 52.4 | 427.5 | 99.0 | 99.0 | 24.0 | 22.5 | | | -1 | AGN | ab | Radio double |
| S155 | ATCDFS_J032850.97-273826.8 | C170 | SWIRE3_J032850.93-273827.1 | 3:28:50.977 | -27:38:26.874 | 0.4 | 78.4 | 58.2 | 40.9 | -1.0 | -1.0 | 99.0 | 99.0 | 23.6 | 22.5 | | | 1 | | | |
| S156 | ATCDFS_J032851.61-280544.6 | C171 | SWIRE3_J032851.63-280544.3 | 3:28:51.610 | -28:05:44.624 | 4.2 | 29.0 | 35.1 | -1.0 | -1.0 | -1.0 | 99.0 | 99.0 | 99.0 | 99.0 | | | 1 | | | |
| S157 | ATCDFS_J032853.28-275401.3 | C172 | SWIRE3_J032853.28-275401.0 | 3:28:53.287 | -27:54:01.353 | 0.3 | 180.7 | 163.9 | 48.2 | -1.0 | -1.0 | 99.0 | 99.0 | 99.0 | 22.2 | | | 1 | | | |
| S158 | ATCDFS_J032854.03-273835.7 | C173 | SWIRE3_J032853.99-273835.4 | 3:28:54.031 | -27:38:35.758 | 0.8 | 42.1 | 25.2 | -1.0 | -1.0 | -1.0 | 99.0 | 99.0 | 22.9 | 21.6 | | | 1 | | | |
| S159 | ATCDFS_J032854.45-271810.0 | C174 | | 3:28:54.452 | -27:18:10.024 | 0.3 | -1.0 | -1.0 | -1.0 | 303.0 | -1.0 | | | | | | | 7 | | | Weak uncatalogued source in IRAC band 1 |

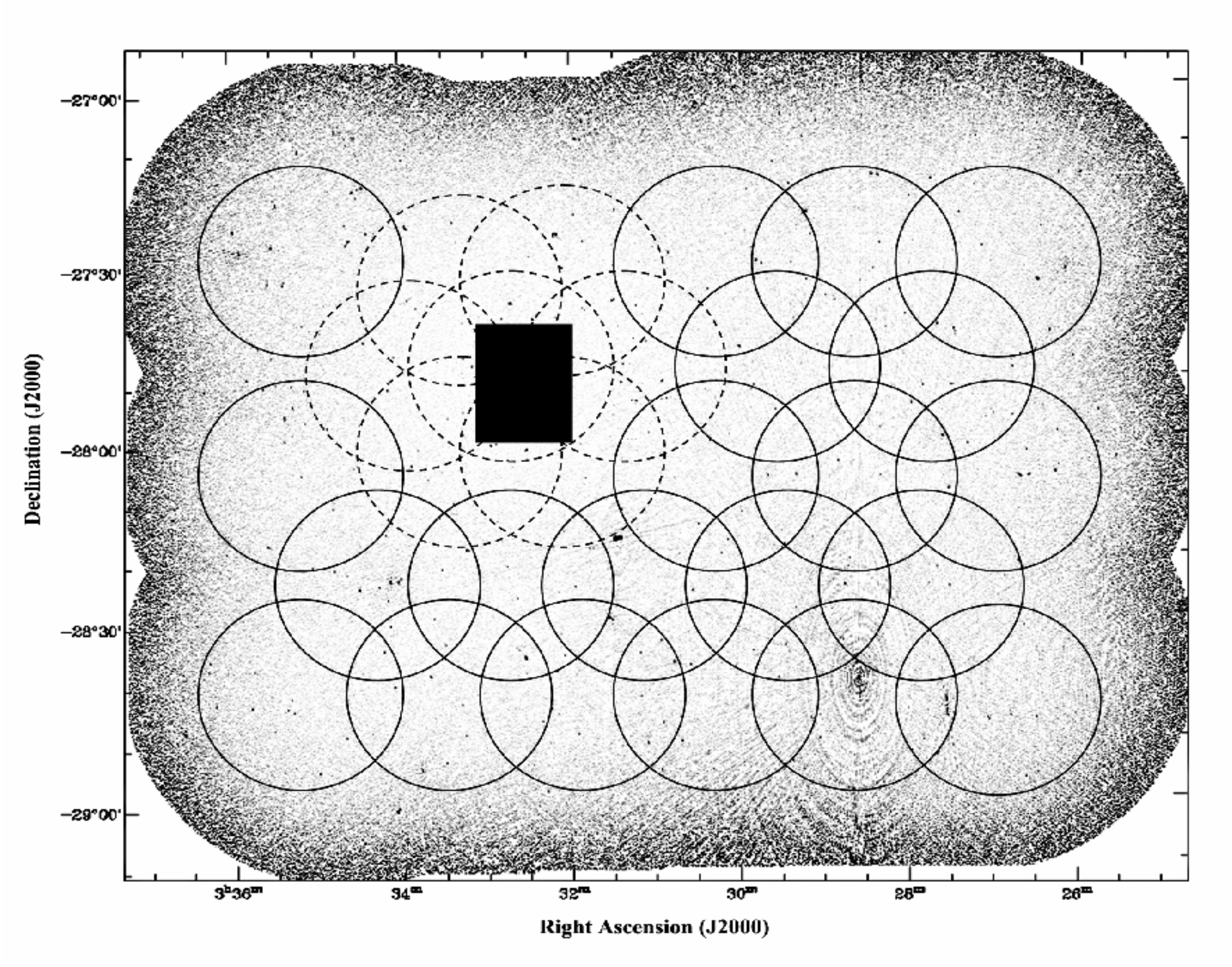

Fig 1: The observed fields, superimposed on the 20cm image. Circles show the half-power beamwidth of the ATCA antennas, centred on the pointing positions used in the observations. Solid circles show areas observed in C1241, dashed circles show fields observed in C1035, and the black square shows the GOODS field.

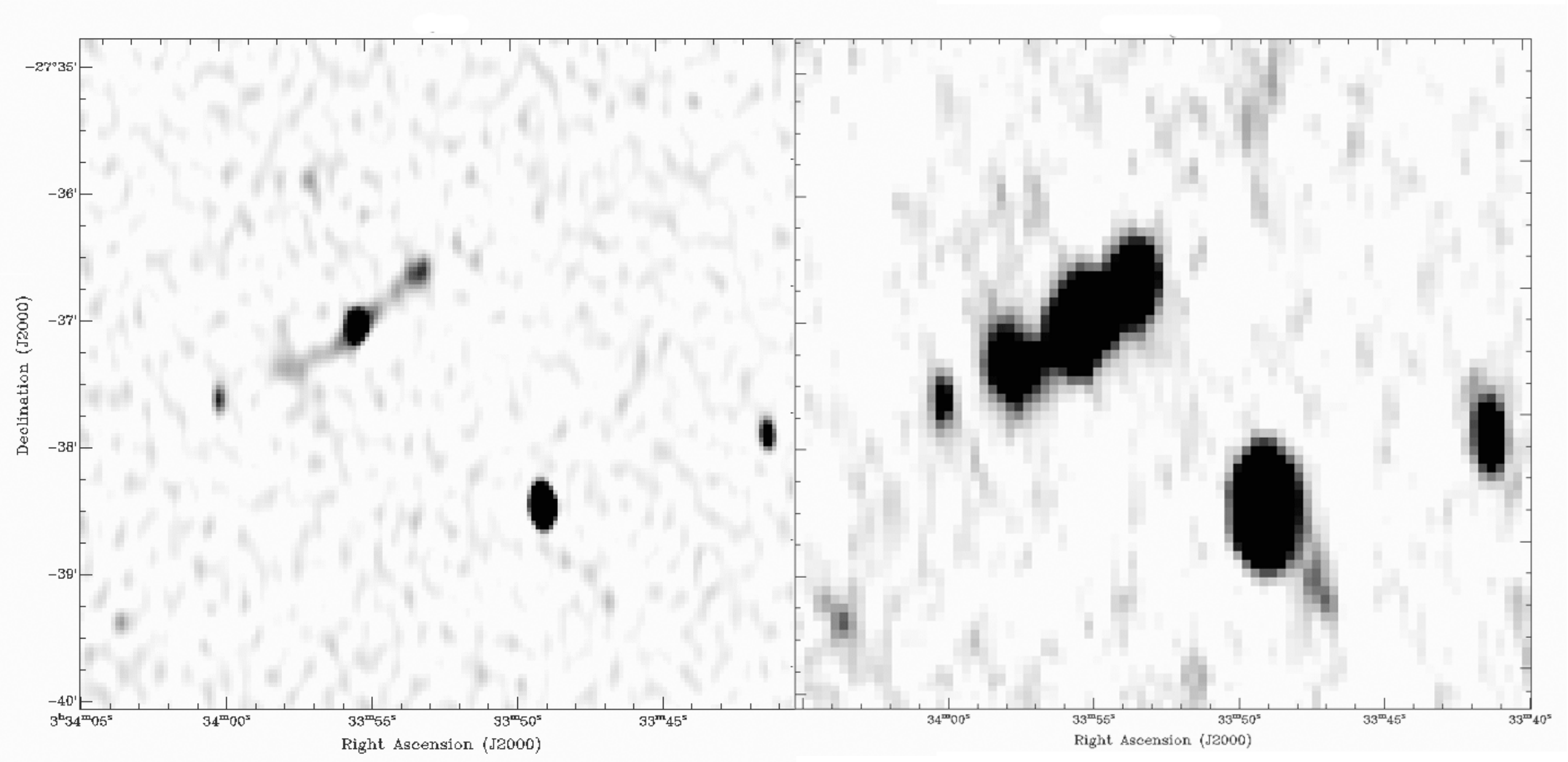

Fig 2: (Left) A representative sample of the radio image used for component identification in this paper, showing a classical triple radio galaxy (C616, C619, C622), together with other single-component galaxies. This image has a beam-size of 11 x 5 arcsec and a local rms noise of 25µJy. (Right) The same image, but made with natural weighting. It has a higher sensitivity (17 µJy rms) but a larger beamsize (28 x 16 arcsec).

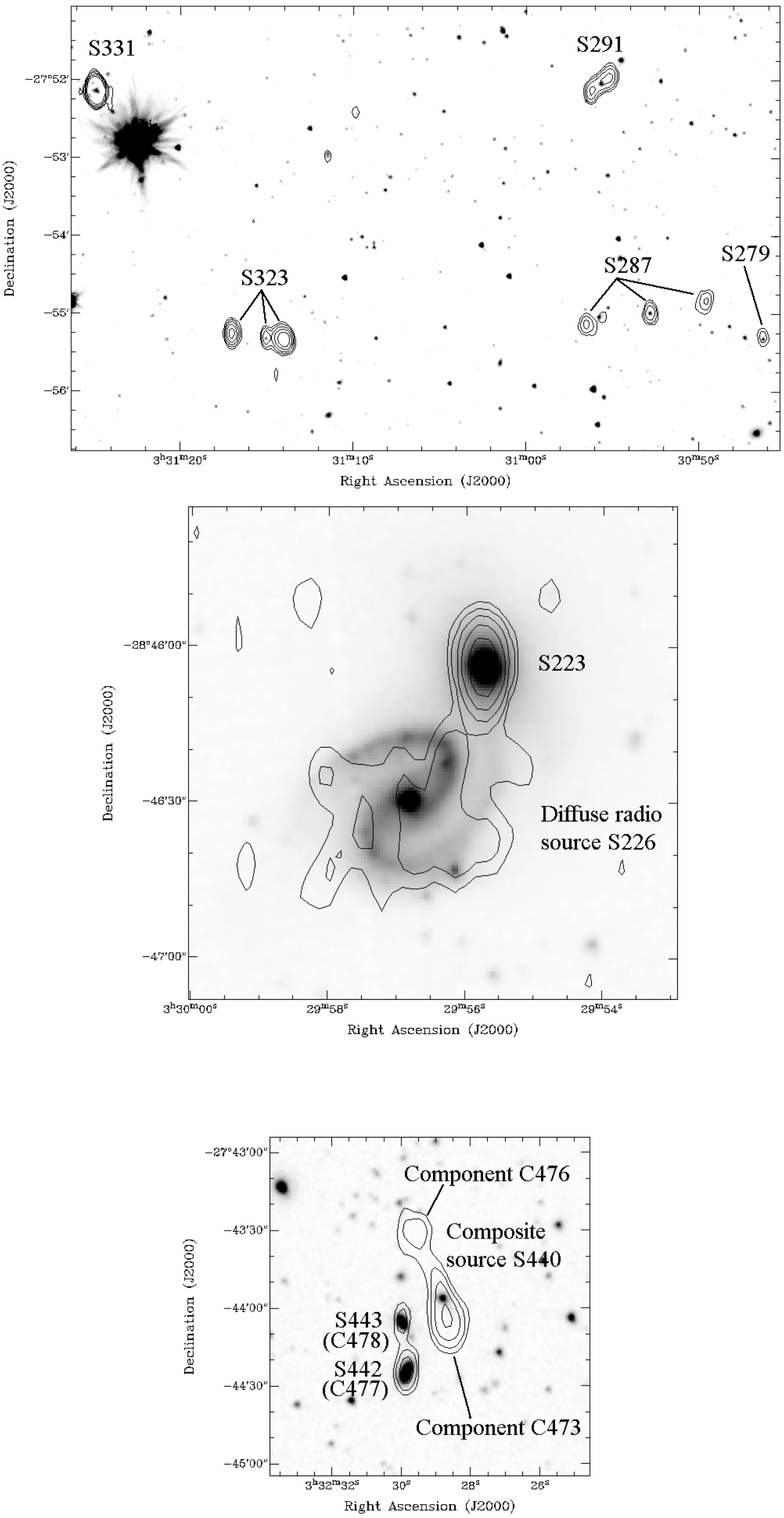

Figure 3. Representative 20 cm images of three regions containing radio sources (contours) overlaid on the 3.6μm Spitzer images (grayscale).

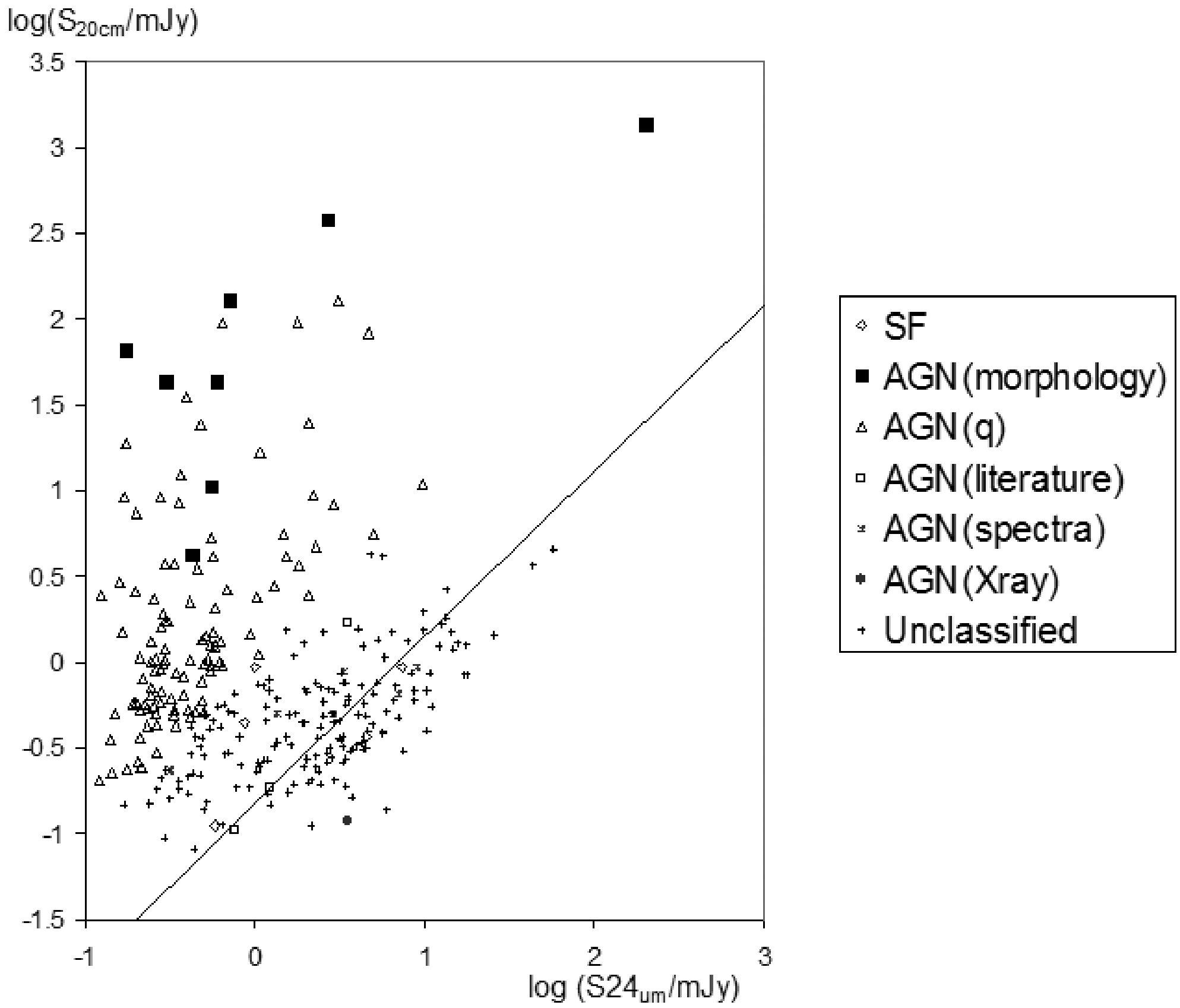

Fig 4: Observed 20cm integrated flux plotted against SWIRE 24µm flux for those sources in our sample which were also detected at 24µm, classified according to the criteria discussed in Section 3.4. The diagonal line shows the radio-FIR correlation $q_{24\mu m} = \log (S_{24\mu m} / S_{20cm}) = 0.84$ suggested by the Appleton et al data.

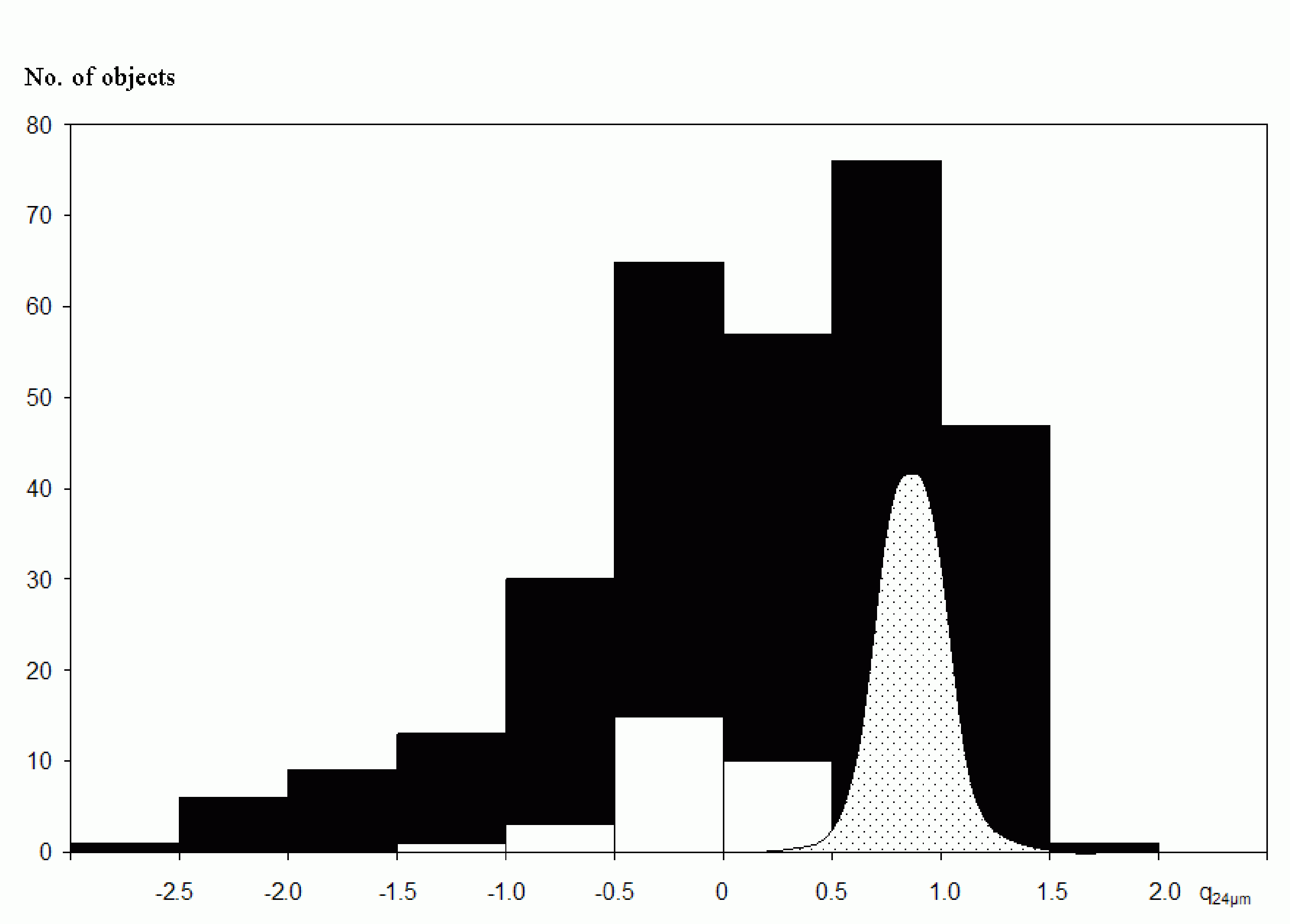

Figure 5: Distribution of values of $q_{24\mu m} = \log (S_{24\mu m}/S_{20cm})$. The black histogram represents the subsample of ATLAS sources for which 24µm fluxes are available. The white histogram represents the upper limits derived by Higdon et al. (2005) for their sample of optically-invisible radio sources, which are believed to be obscured AGN. The Gaussian curve shows the approximate distribution of the sample of Appleton et al. (2004), which is dominated by star-forming galaxies.

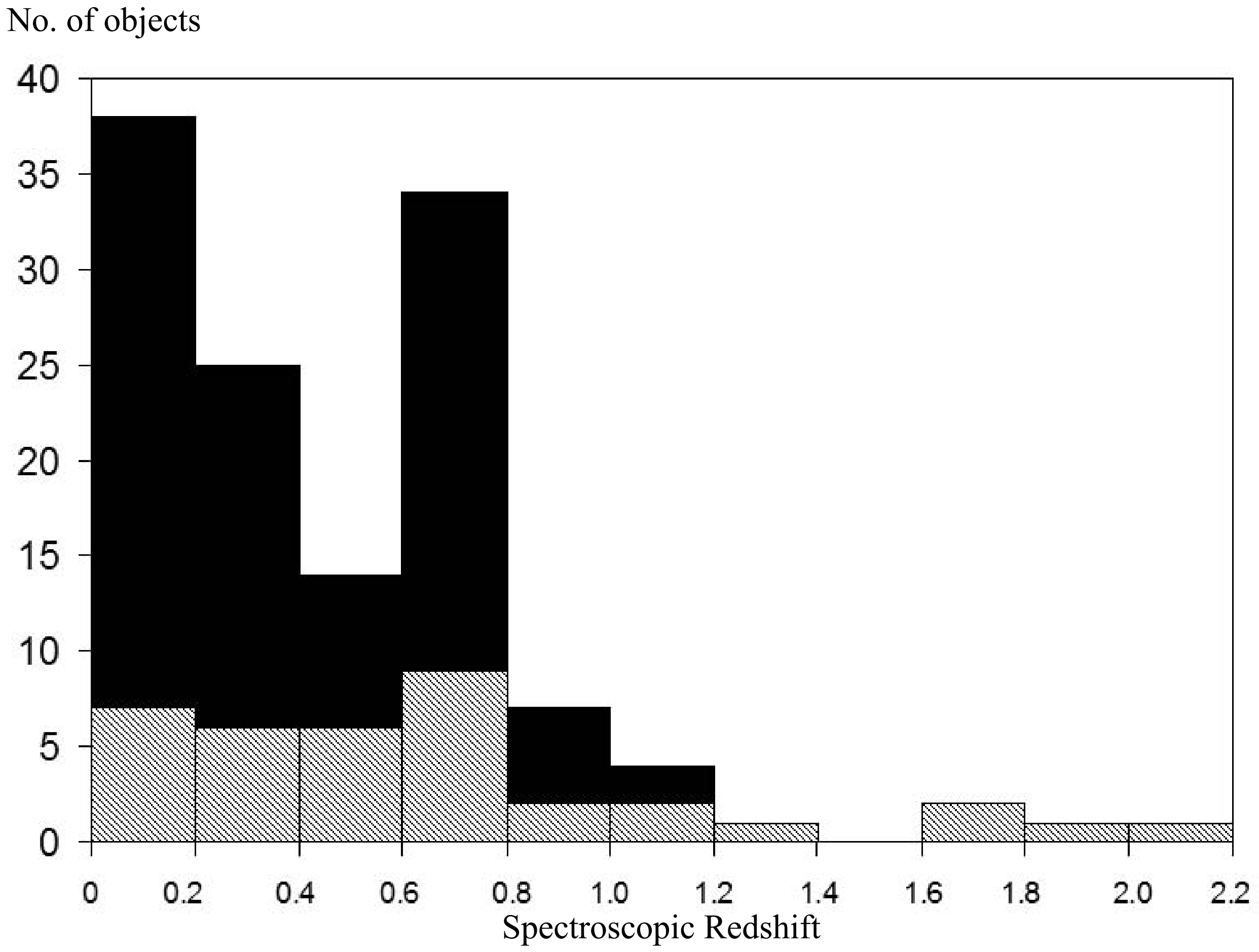

Fig 6: Histogram of spectroscopic redshifts for our sample. The upper curve (black area) shows the total for all objects, whilst the lower curve (shaded area) shows objects classified as AGN.

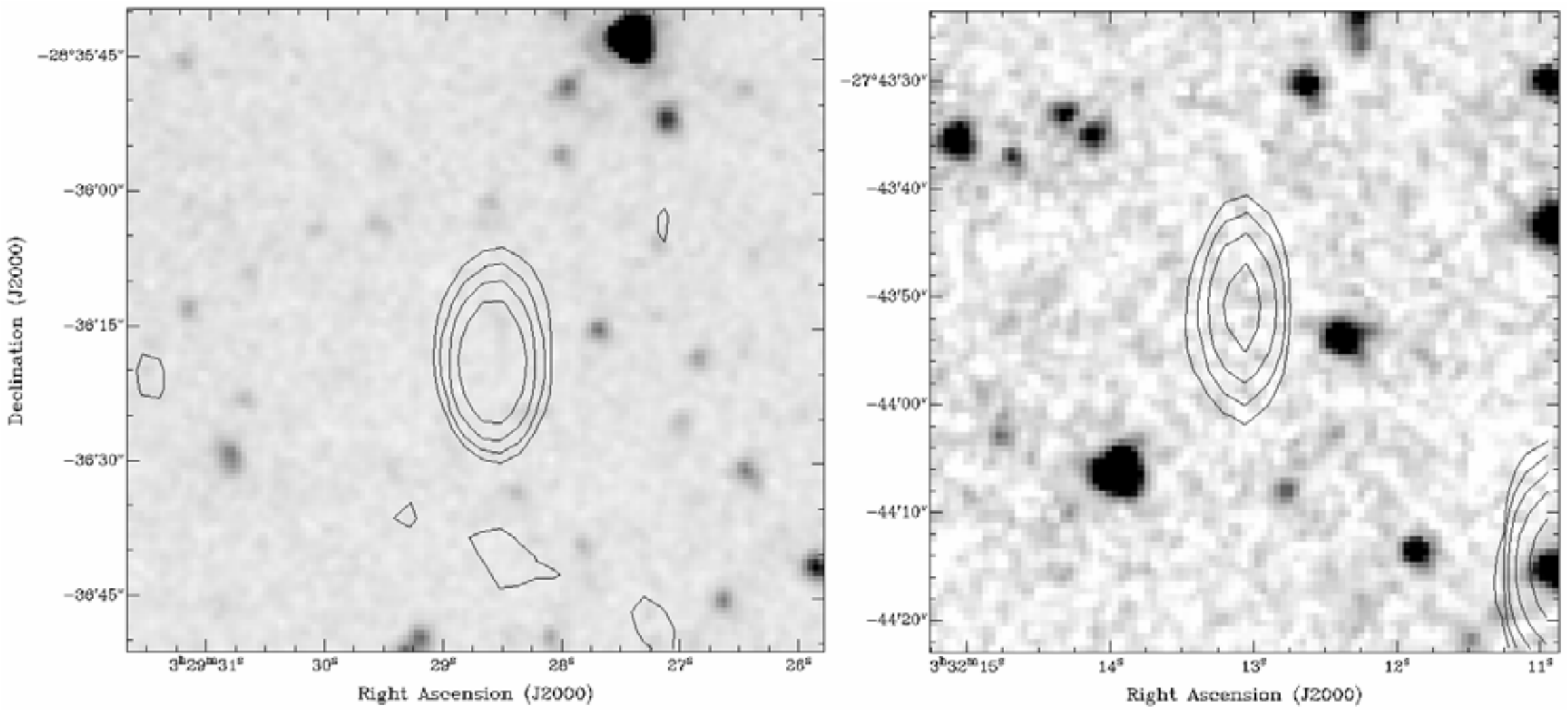

Fig 7. Two infrared-faint radio sources, both of which are bright (~6 mJy) at 20cm, but have no known infrared, optical, or X-ray counterpart.

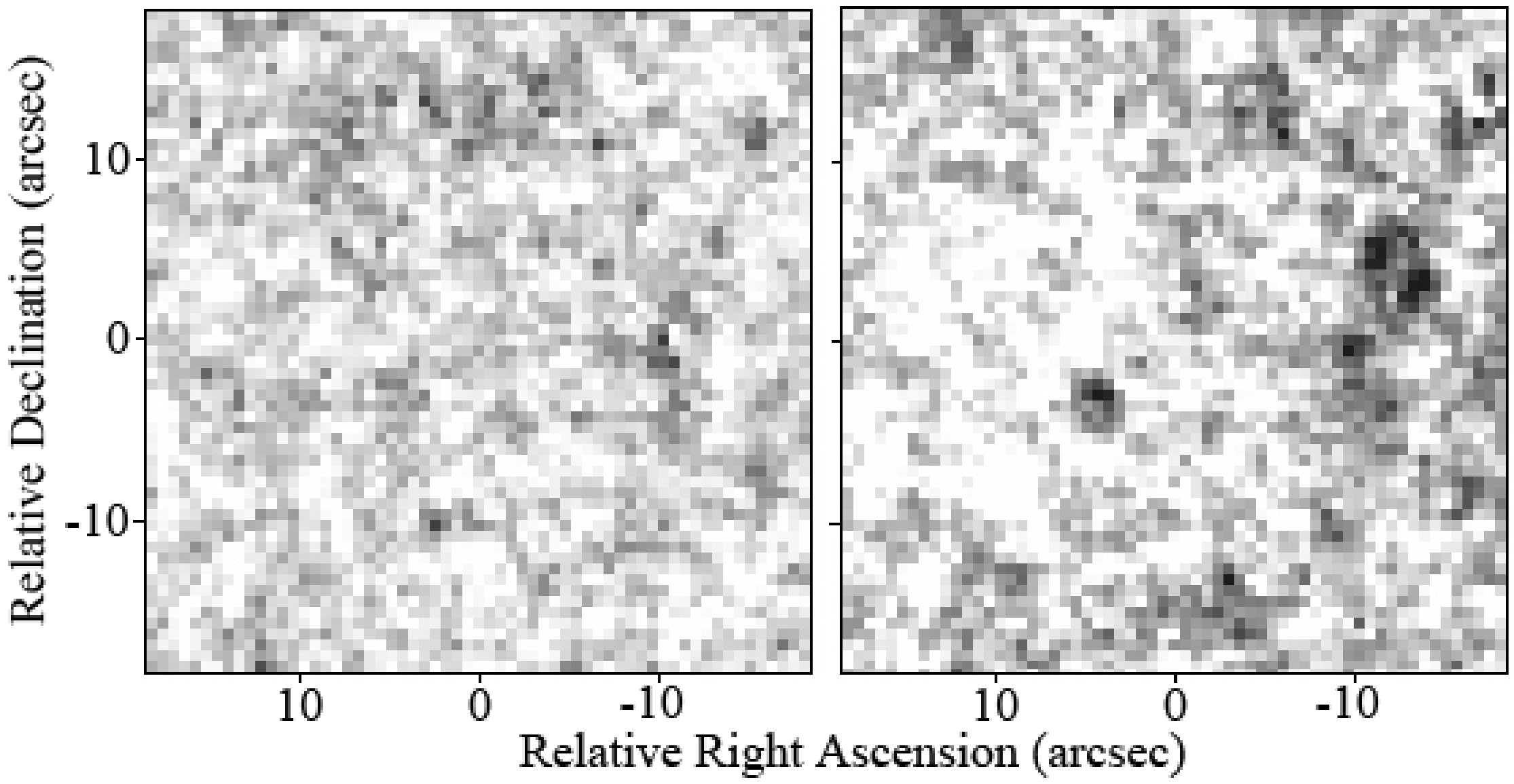

Fig 8. (Left) Stacked 3.6 µm IRAC image of all 22 radio sources for which there is no infrared counterpart. (Right) Stacked 3.6 µm IRAC image of the eight brightest radio sources for which there is no infrared counterpart. In both cases, the position of the radio source is at the centre.